\documentclass[aps, prx, notitlepage, citeautoscript, superscriptaddress, reprint]{revtex4-2}
\usepackage{subcaption}
\usepackage{hyperref}
\usepackage{graphicx}
\usepackage{multirow}
\usepackage{siunitx}
\usepackage{amssymb}
\usepackage{hyperref}

\begin{document}

\preprint{APS/123-QED}

\title{Multi-qubit Dynamical Decoupling for Enhanced Crosstalk Suppression}

\author{Siyuan Niu}
\email{siyuan.niu@lirmm.fr, siyuanniu@lbl.gov}
 \affiliation{%
  LIRMM, University of Montpellier, France
}

\author{Aida Todri-Sanial}
 \affiliation{%
  LIRMM, University of Montpellier, CNRS, France
}
\affiliation{Eindhoven University of Technology, Eindhoven, Netherlands.}

\author{Nicholas T. Bronn}
\email{ntbronn@us.ibm.com}
 \affiliation{%
  IBM Quantum, IBM T. J. Watson Research Center, Yorktown Heights, NY 10598, United States of America
}





\date{\today}

\begin{abstract}

Dynamical decoupling (DD) is one of the simplest error suppression methods, aiming to enhance the coherence of qubits in open quantum systems. Moreover, DD has demonstrated effectiveness in reducing coherent crosstalk, one major error source in near-term quantum hardware, which manifests from two types of interactions. Static crosstalk exists in various hardware platforms, including superconductor and semiconductor qubits, by virtue of always-on qubit-qubit coupling. Additionally, driven crosstalk may occur as an unwanted drive term due to leakage from driven gates on other qubits. Here we explore a novel staggered DD protocol tailored for multi-qubit systems that suppresses the decoherence error and both types of coherent crosstalk. We develop two experimental setups -- an ``idle-idle'' experiment in which two pairs of qubits undergo free evolution simultaneously and a ``driven-idle'' experiment in which one pair is continuously driven during the free evolution of the other pair. These experiments are performed on an IBM Quantum superconducting processor and demonstrate the significant impact of the staggered DD protocol in suppressing both types of coherent crosstalk. When compared to the standard DD sequences from state-of-the-art methodologies with the application of X2 sequences, our staggered DD protocol enhances circuit fidelity by 19.7\% and 8.5\%, respectively, in addressing these two crosstalk types.


\end{abstract}

\maketitle


\section{\label{sec:1}Introduction}

The inherent noise in near-term quantum computers hinders the demonstration of practical quantum advantage. Recently, a variety of quantum error mitigation techniques have been developed to tackle this challenge, such as zero noise extrapolation (ZNE)~\cite{li2017efficient, temme2017error}, Clifford data regression (CDR)~\cite{czarnik2021error}, and probabilistic error cancellation (PEC)~\cite{van2023probabilistic}, which require substantial overhead in quantum resources (such as sampling overhead) in order to model the noise and mitigate the resulting error. Specifically, ZNE, CDR and PEC incur linear, polynomial, and exponential sampling overheads, respectively. Among these methods, only PEC provides an accurate and bias-free estimate. In contrast, quantum error suppression techniques, such as randomized compiling~\cite{Wallman2016}, pulse-scaling for Trotterized circuits~\cite{Stenger2021, earnest2021pulse}, and dynamical decoupling (DD)~\cite{viola1998dynamical, viola1999dynamical, biercuk2009optimized}, offer a reduction in error with no additional quantum resources. However, randomized compiling necessitates running circuits multiple times and averaging the results for effective error suppression, which may introduce latency due to the increased bandwidth when implementing on cloud-based quantum computers. Pulse-scaling does not require additional resources if the default pulses are parametrically scaled, otherwise the overhead of additional calibration experiments must be considered. DD aims to reduce the decoherence error by decoupling the interaction between the quantum system and the environment. This process involves inserting a sequence of gates that compose the identity during the idle period when the qubit is not being used for computation. This entails very little {\it classical} overhead, as it only requires identifying the idle times and inserting sequences with a pattern of delays between the gates.
It makes DD one of the simplest error suppression methods, as it involves no additional sampling or experiment execution overhead. Numerous DD sequences have been introduced over decades~\cite{meiboom1958modified, viola1999dynamical, zhang2014protected, uhrig2007keeping, genov2017arbitrarily}, leading to performance improvements on various modalities of quantum computers, including superconducting circuits~\cite{niu2022effects,pokharel2023demonstration, souza2021process,ezzell2022dynamical, tripathi2022suppression,tong2024empirical}, trapped ions~\cite{bermudez2012robust, morong2023engineering}, neutral atoms~\cite{bluvstein2022quantum}, and semiconductors~\cite{medford2012scaling}. DD achieves these performance improvements by reducing not only decoherence error but also crosstalk, which is a major error source in these quantum technologies~\cite{parrado2021crosstalk, Sheldon2016, auger2017blueprint, buterakos2018crosstalk}. These studies typically focus on a single qubit and its environment but quantum computers consist of many qubits coupled to each
other. Notable exceptions~\cite{zhou2023quantum, tripathi2022suppression,shirizly2024dissipative,mundada2023experimental} explore static $ZZ$ crosstalk in multi-qubit systems. They propose dynamical decoupling (DD) protocols to protect the quantum systems from crosstalk and enhance the gate fidelity, similar in pattern to those we consider.



Here, we investigate a different multi-qubit DD strategy called {\it staggered} DD that consists of a standard DD sequence with symmetric free evolution times on one qubit with the same sequence on an adjacent qubit interleaved in time such that the gates occur in the middle of the free evolution times of the original sequence. In addition to suppressing dephasing and single-qubit coherent errors, we show that staggered DD can suppress undesired two-qubit interactions by inverting their direction of rotation in the two-qubit Hilbert space during each of the half-idle times of the gate interleaving. We first provide some theoretical insight for this method, by placing it in context with the usual treatment of DD for general decoherence. We show that this can suppress crosstalk from unwanted $ZZ$ interactions, which can arise both from static coupling between transmon superconducting qubits and driven crosstalk from cross resonance interaction. We then conduct experiments on an IBM quantum computer to explore both static and driven crosstalk. Even though the standard DD sequence has already shown to reduce the decoherence error and crosstalk, the experimental results demonstrate that staggered DD performs better than standard (unstaggered) sequences applied na\"\i vely to all qubits.

\section{Theoretical analysis}\label{sec:theory}

The general Hamiltonian representation for an open quantum system is
\begin{equation}
    H = H_S + H_B + H_{SB}
\end{equation}
where $H_S$ and $H_B$ correspond to the system and bath Hamiltonian respectively, while $H_{SB}$ signifies the interaction between the system and the bath. For a single qubit, the general system-bath interaction (taking $\hbar \equiv 1$ throughout) 
\begin{equation}
H_{SB} = \sum\limits_{\alpha = x,y,z} \sigma^{\alpha} \otimes B_{\alpha}
\end{equation}
includes both {\it coherent} and {\it incoherent} single-qubit errors, where $\sigma^{\alpha}$ ($B_\alpha$) are the Pauli (bath) operators. The objective of dynamical decoupling is to suppress the unwanted error terms of the total Hamiltonian to a certain order. Taking the simplest DD sequence -- X2 as an example, a standard X2 sequence is composed of a repetition of 
\begin{eqnarray}
{\rm q}: f_{\tau/2} - \sigma^x - f_{\tau} - \sigma^x - f_{\tau/2},
\end{eqnarray}
for each qubit $q$, with $\tau$ being the interval between pulses for free evolution $f$, throughout a total idle period of $2\tau$. The effect of the first $\sigma^x$ operation transforms the system-bath Hamiltonian to 
\begin{eqnarray}
H_{SB}' & = & (\sigma^x \otimes I_{B}) H_{SB} (\sigma^x \otimes I_{B}),
\end{eqnarray}
where $I_B$ represents the identity operator on the bath, so that the free evolution during $f_\tau$ is
\begin{eqnarray}
U_{SB}' = e^{-i\tau(\sigma^x \otimes B_x - \sigma^y \otimes B_y - \sigma^z \otimes B_z)}.
\end{eqnarray}

\noindent
Ignoring the $\sigma^x$ term, $H_{SB}' = -H_{SB}$ (i.e., $U_{SB}' = U_{SB}^\dagger$), therefore the $Y$ and $Z$ terms are cancelled to the first order based on the application of the Baker-Campbell-Hausdorff (BCH) formula to the system-bath interaction over the entire evolution time of $2\tau$. While the X2 sequence leaves the $X$ term unchanged, it can be removed by more complicated DD sequences such as XY4. 

Transmons are a common style of superconducting qubit, where the quantum infomation is encoded in the frequencies of oscillations of Cooper pairs through a Josephson junction~\cite{Koch2007}. Two fixed-frequency transmons with fixed coupling may be modelled as Duffing oscillators
\begin{eqnarray}
H_{\rm Duff} & = & \sum_{i \in \{0,1\}} \left( \omega_i \hat{a}_i^\dagger \hat{a}_i + \frac{\delta_i}{2} \hat{a}_i^\dagger \hat{a_i} \left( \hat{a}_i^\dagger \hat{a}_i - 1 \right) \right) \\
  & & + J\left(\hat{a}_0^\dagger + \hat{a}_0 \right) \left( \hat{a}_1^\dagger + \hat{a}_1 \right) \nonumber
\end{eqnarray}
where $\omega_i$ and $\delta_i$ are the bare transmon frequencies and anharmonicities, respectively, $\hat{a}_i^\dagger$ ($\hat{a}_i$) are the creation (annihilation) operators for transmon $i$ and $J$ is the coupling strength. This coupling dresses the transmon frequencies and introduces an always-on static $ZZ$ crosstalk interaction with strength
\begin{eqnarray}
\nu_{ZZ} & = & \left(\omega_{11} - \omega_{01} \right) - \left( \omega_{10} - \omega_{00} \right) \\
  & \approx & \frac{2 J^2(\delta_0 + \delta_1)}{(\delta_1 - \omega_{10} + \omega_{01}) (\delta_0 + \omega_{10} - \omega_{01})} \nonumber
\end{eqnarray}
in the perturbative limit where $\omega_{ij}$ refers to the energy of the two-qubit state $|ij\rangle$~\cite{Wei2022}.  Neglecting the system-bath interaction, this generates a coherent rotation in the two-qubit subspace of $ZZ(\theta)$ with $\theta = \nu_{ZZ} t$ where $t$ is the free evolution time. Note that when $\theta$ is unspecified, the Pauli operator is assumed (i.e., $\theta=\pi$). This static interaction intrinsically limits quantum volume~\cite{Jurcevic2020}, performance of quantum error correction codes~\cite{Takita2017}, and may prevent scalability of quantum processors~\cite{Berke2022}.

\begin{figure*}[!t]
\centering
\includegraphics[scale=0.5]{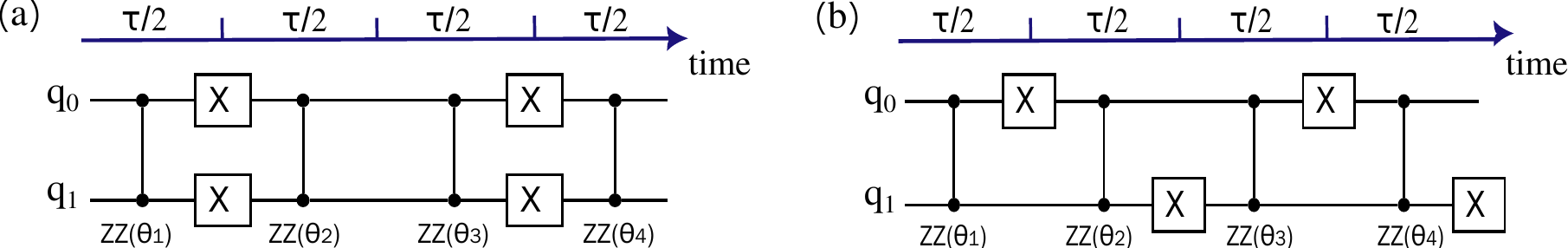}

\caption{A quantum circuit evolves an idle period of $2\tau$ and suffers from crosstalk $ZZ(\theta)$, with (a) standard X2 sequence inserted, and (b) staggered X2 sequence inserted.}
\label{timeline}
\end{figure*}

Applying the X2 DD sequence similarly across multiple qubits, the free evolution for each qubit is identical, as $XX$ terms commute with the $ZZ$ crosstalk (we adopt Pauli-string notation, i.e. $XX \equiv \sigma_0^x \otimes \sigma_1^x$, where the position in the string indicates which qubit the Pauli acts on),
\begin{equation}
XX \cdot ZZ(\theta) \cdot XX = ZZ(\theta)
\end{equation}
thus, the crosstalk accumulates across all the idle times.
\begin{eqnarray}
e^{-i ZZ \theta_4/2} XX e^{-i ZZ (\theta_3 + \theta_2)/2} XX e^{-i ZZ \theta_1/2} \\
 = \exp\{-i ZZ (\theta_1 + \theta_2 + \theta_3 + \theta_4)/2\} \nonumber
\end{eqnarray}
If instead, only a single gate is applied to each qubit at a time, the crosstalk can be inverted,
\begin{equation}
XI \cdot ZZ(\theta) \cdot XI = IX \cdot ZZ(\theta) \cdot IX = ZZ(-\theta)
\end{equation}
The goal of the staggered X2 sequence, in which the DD sequences differ for two qubits $(q_0, q_1)$, is to cancel the accumulated two-qubit $ZZ$ rotation by staggering the timing of the gates in the DD sequences. Specifically, the staggered X2 sequence
\begin{eqnarray}
& {\rm q}_0: & f_{\tau/2} - X - f_{\tau} - X - f_{\tau/2} \\
& {\rm q}_1: & f_{\tau} - X - f_{\tau} - X \nonumber
\end{eqnarray}
creates the free evolution
\begin{eqnarray}
IX e^{-i ZZ \theta_4/2} XI e^{-i ZZ \theta_3/2} IX e^{-i ZZ\theta_2/2} XI e^{-i ZZ \theta_1/2} \\
 = \exp\{-i ZZ (\theta_1 - \theta_2 + \theta_3 - \theta_4)/2\} \nonumber
\end{eqnarray}
in which the accumulated phases cancel because each $\theta_i$ is an equal rotation over the free evolution time $\tau/2$. An illustration of applying the standard or staggered X2 sequence to a two-qubit circuit is shown in figure~\ref{timeline}. We can also invert the staggered DD sequences between the control and target qubits to the same effect. This staggered technique can likewise be extended to other DD sequences, such as 
\begin{equation}
\mathrm{XY4} = X - Y - X - Y
\end{equation}
and
\begin{equation}
\mathrm{XY8} = X - Y - X - Y - Y - X - Y - X,
\end{equation}
where explicit references to the free evolutions $f_\tau$ have been omitted for simplicity.  Noisy simulations illustrating the cancellation of $ZZ$ crosstalk via the use of staggered DD are shown in~\ref{app:sim}. 
 In the following, we study the differences between staggered and standard DD sequences of different orders (X2, XY4, XY8) by inserting idle times (representing free evolution of time $2\tau$) between Clifford gates in two-qubit randomized benchmarking (RB) sequences~\cite{Magesan2011, Magesan2012}. These RB sequences are performed individually and in conjunction with operations on an adjacent pair of qubits, demonstrating that staggered DD consistently suppresses coherent error more effectively, whether it is caused by static transmon-transmon higher-level coupling or cross-resonance-driven crosstalk.

\begin{figure}
    \centering
    \includegraphics[scale=0.8]{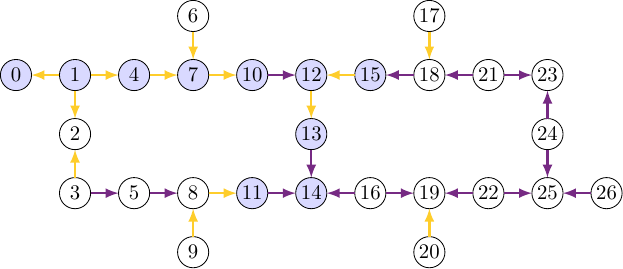}
    \caption{Qubit coupling graph of 27-qubit {\it ibm\_cairo} backend. The selected qubits for experiments are highlighted in blue circles. The qubit connections follow the native directions. ECR and DCX types of $\texttt{CX}$ gates are represented by yellow and purple colors respectively.}
    \label{fig:coupling}
\end{figure}
\begin{figure*}[t]
\centering
\includegraphics[scale=0.47]{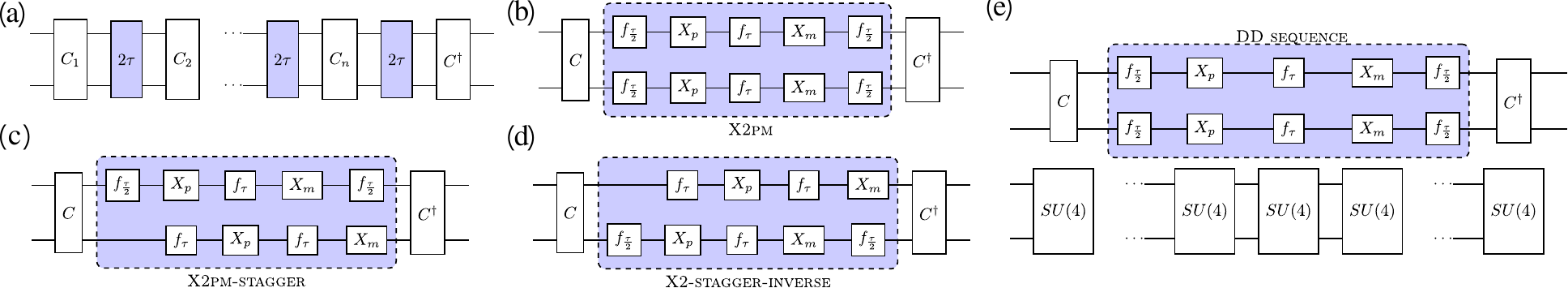}
\caption{The RB circuit for a qubit pair is shown in (a). We insert X2{pm} DD sequence in (b), staggered X2pm sequence in (c) and inverse of staggered X2pm in (d). The circuit for driven-idle experiment composed of RB and SU(4) is shown in (e).}
\label{SRB_cir}
\end{figure*}
\section{Results}\label{sec:results}
\subsection{Experimental Protocol}


We select a subset of ten qubits from the 27-qubit {\it ibm\_cairo} backend (the coupling graph is shown in figure~\ref{fig:coupling}), which includes controlled-NOT $\texttt{CX}$ pairs generated by both echoed cross resonance (ECR)~\cite{Chow2011, Sheldon2016} and direct CX (DCX)~\cite{Jurcevic2020}. We conduct experiments on two qubit pairs to study the crosstalk between them. The two qubit pairs are chosen to be physically connected for two reasons: (1) Connected qubits within each pair can introduce static crosstalk.
(2) Operations on one qubit pair can induce more driven crosstalk to its neighboring qubit pair~\cite{Murali2020, Wilson2020, niu2021analyzing}. We present the RB experiments in two experimental settings: (i) {\it idle-idle} in which the two pairs of qubits are simultaneously idle during their simultaneous RB sequences (aside from the relevant DD sequences) in Section~\ref{sec:idle-idle}, and (ii) {\it idle-driven} in which the fidelity of the vulnerable pair of qubits is measured after the RB sequence (with varying DD sequence)  while the other pair of qubits is driven by a random sequence of two-qubit gates (SU(4)) in Section~\ref{sec:cross-idle}. The two-qubit \emph{isolated} RB circuit sets a baseline for the two experiments for comparison,  where the $ZZ$ interaction between qubits in the pair is suppressed by calibrating rotary echo pulses on the target qubits~\cite{sundaresan2020reducing}. This is compared to the same RB circuit but while a neighboring qubit pair is active, effectively exposing the RB circuit to static and/or driven crosstalk (the effect of static crosstalk in which the neighboring qubits are in the ground state has been calibrated out by the backend). The idle-idle and idle-driven experiments expand to involve two qubit pairs, totaling four qubits. Standard and staggered DD sequences are inserted during the idle periods. Specifically, in the idle-idle experiment, DD are applied to both qubit pairs, whereas in the second experiment, DD is only applied to the first qubit pair with RB sequences. These experiments highlight the difference in performance between standard and staggered DD sequences when the qubits are exposed to both stochastic dephasing and coherent crosstalk arising from static $ZZ$ coupling and cross-resonance driven $ZZ$. The quantum circuits were generated and executed using the \emph{Qiskit Experiments} framework~\cite{Kanazawa2023}. 

In the following experiments, the two-qubit RB circuits consist of a fixed number of Clifford gates (Clifford sequence) separated by varying lengths of the idle period of $2\tau$ ranging from \SI{284.4}{\nano\second} to \SI{3128.9}{\nano\second} as shown in figure~\ref{SRB_cir}a. Note that these idle durations are chosen as multiples of the sampling rate of the pulse waveform generator $dt=0.2\overline{2}$\,\si{\nano\second} on \textit{ibm\_cairo}. The same Clifford sequence with $n=8$ gates and an additional Clifford gate $C^\dagger$ is applied to invert all the previous operations to the initial state. This setting maintains a high initial fidelity ($\gtrsim 0.8$) at the minimum idle time.
Note that the circuit depth remains constant throughout the experiment, with the only varying parameter being the length of the delay that separates the sets of Clifford gates. 


We choose two qubit pairs $(q_{11}, q_{14})$ and $(q_{12}, q_{13})$ for study, that are particularly vulnerable to crosstalk from operations on neighboring qubit pair and demonstrates stable output, as shown by the fidelity differences between figure~\ref{fig:idle_res_X2}a and figure~\ref{fig:idle_res_X2}b. Data for the rest of the qubit pairs from the ten selected qubits are detailed in~\ref{app:1}.
For idle-idle experiment, we apply exactly the same RB circuit simultaneously to the adjacent pair of qubits. While the duration of the idle period remains the same for both qubit pairs, the starting points of these idle periods differ slightly due to the differences in duration of the two neighboring $\texttt{CX}$ pairs. We use the \emph{As Late As Possible} policy to schedule the circuits.
For the idle-driven experiment, while one qubit pair undergoes the RB circuit, its adjacent pair is subjected to a circuit composed of different random SU(4) blocks, which continuously drives crosstalk during the RB experiment, as shown in figure~\ref{SRB_cir}e. The number of SU(4)s in the circuit is determined to ensure it has the same duration as the RB circuit. This experimental arrangement ensures the simultaneous execution of both circuits. Since no idle time is inserted between SU(4)s, during the idle periods of the RB circuit, the neighboring qubits in the SU(4) circuit remain active thereby contributing to driven crosstalk. In the two experiments, the $\texttt{CX}$ gate is aligned with the native direction of either ECR or DCX gates, determined in calibration to that of higher fidelity.

We measure the probability of return to the initial state (as it is related to the randomized benchmarking protocol), to gauge the fidelity changes across time.  Moreover, we numerically calculate the time average fidelity~\cite{zhou2023quantum, ezzell2022dynamical}, 
\begin{equation}
    F_{\rm avg} (t) = \frac{1}{T} \int_{0}^{T} dt \frac{F(t)}{F(0)}
    \label{eq:fid}
\end{equation}
to indicate the overall fidelity change over time, where $F(t)$ represents the probability of return at time $t$.

 During the idle period, DD sequences implemented in different ways at the pulse and timing level are inserted as depicted in figure~\ref{SRB_cir}b-d. Standard DD sequences have symmetric timing on both qubits, and they occur simultaneously (see figure~\ref{SRB_cir}b for a representation of the X2pm sequence). For staggered sequences, the sequence on one qubit is symmetric (as in the standard case) and the other is offset such that each staggered pulse occurs halfway between the symmetric pulses on the other qubit (figure~\ref{SRB_cir}c). The ``inverse'' staggered sequence switches the role of symmetric and staggered sequences on the qubits (figure~\ref{SRB_cir}d). While each sequence is nominally identical from the perspective of unitary operations (composing to the identity), there are control variations at the pulse level~\cite{ezzell2022dynamical} that may affect their performance. We consider the default \texttt{X} gate to be a $+\pi$ rotation around the $x$-axis of the Bloch sphere (also denoted \texttt{Xp}) and \texttt{Xm} a $-\pi$ rotation with negative amplitude at the pulse level. These variations consisting of alternating rotations are denoted with \texttt{pm} after the sequence name. For example, the X2pm sequence consists of the gates \texttt{Xp-Xm}.


\subsection{Idle-idle Experiment - Simultaneous Randomized Benchmarking}\label{sec:idle-idle}

We evaluate the performances of standard DD sequences, including X2/X2{pm}, XY4/XY4{pm}, XY8/XY8{pm}, and their staggered and inverse of staggered variants - 18 sequences in total. We then compare these sequences to a baseline, in which the qubits remain free evolution during the delay, with no DD sequence inserted as shown in~figure~\ref{SRB_cir}a. 

\begin{figure}[h]
\begin{subfigure}{\columnwidth}
    \centering
    \includegraphics[scale=0.5]{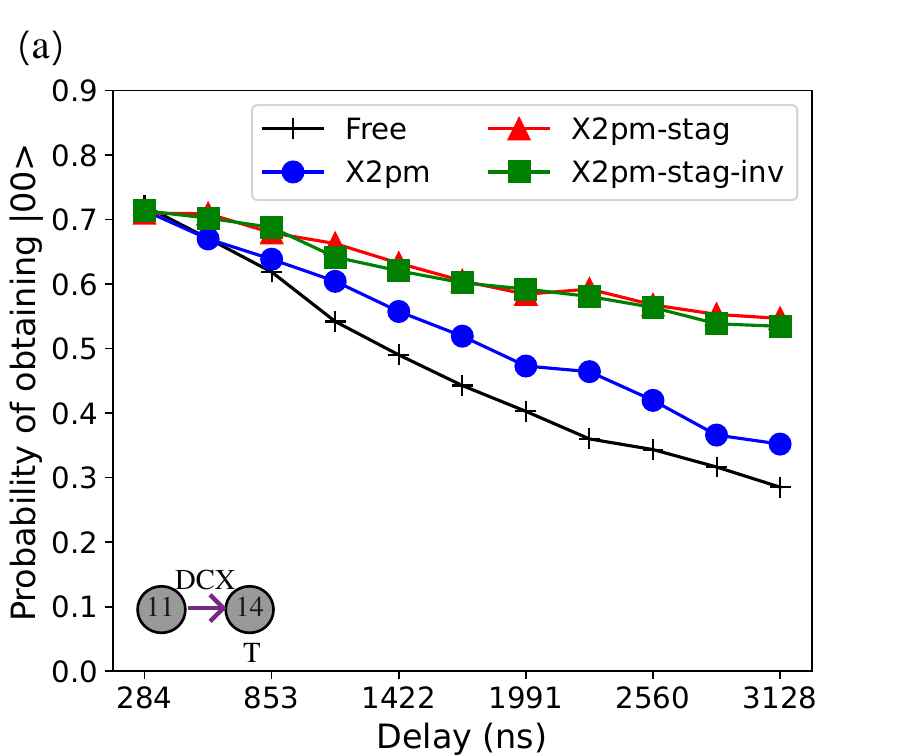}
\end{subfigure}
\begin{subfigure}{\columnwidth}
\centering
\includegraphics[scale=0.5]{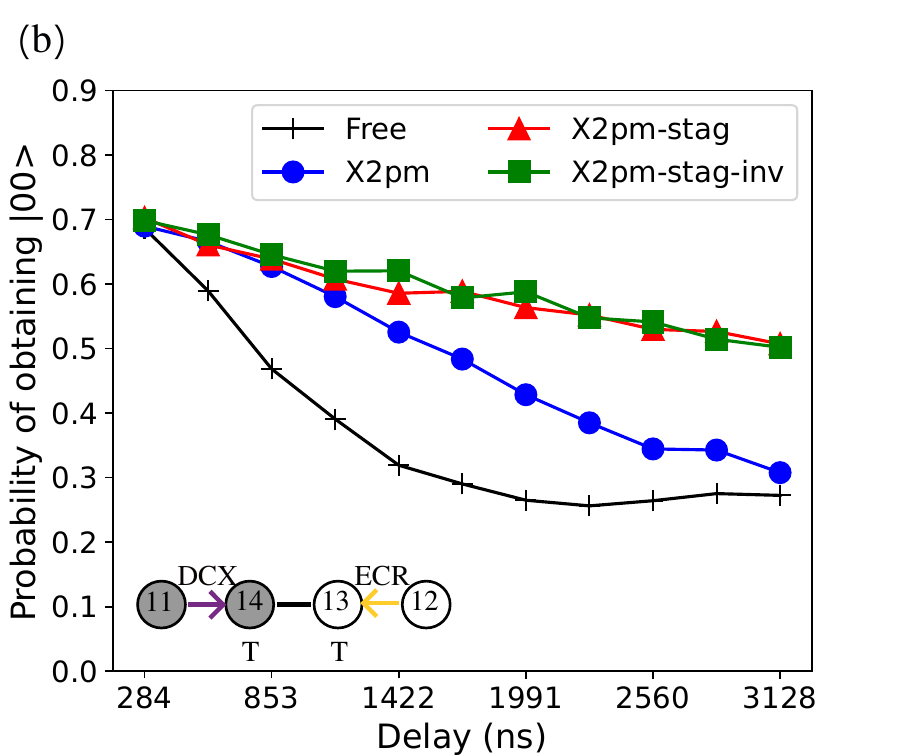}

\end{subfigure}\\
\caption{Idle-idle experiment results on qubit pairs $(q_{11}, q_{14})$ and $(q_{12}, q_{13})$ when comparing the baseline (free evolution) with different DD sequences: X2pm, staggered X2pm, and inverse of staggered X2pm. (a) The probability of yielding $|00\rangle$ for $(q_{11}, q_{14})$ when RB is applied to this qubit pair in isolation. (b) The same probability when RB is applied to both qubit pairs simultaneously (SRB experiment). The significant difference in fidelity during free evolution between the RB and SRB experiments indicates additional crosstalk generated by the neighboring qubit pair. Note that ``T" indicates the qubit serves as the target for the $\texttt{CX}$ gate applied to the qubit pair. The results correspond to the gray qubit pair.}

\label{fig:idle_res_X2}
\end{figure}

The RB and simultaneous RB (SRB) results for the qubit pair ($q_{11}, q_{14}$) are illustrated in figure~\ref{fig:idle_res_X2}, showcasing the comparison between various X2 sequences (X2{pm}, X2{pm}-stag, X2{pm}-stag-inv) against the free evolution. In the SRB experiment, it involves the adjacent qubit pair of ($q_{11}, q_{14}$), which is ($q_{12}$, $q_{13}$), with $q_{14}$ and $q_{13}$ being the target qubits in their respective pairs and neighboring each other. The figure  shows the connectivity of the two qubit pairs, detailing the type of $\texttt{CX}$ gate used (ECR or DCX) and identifying whether a qubit acts as a control (C) or target (T) for the neighboring qubit connections between pairs. The qubit pair ($q_{11}, q_{14}$) is highlighted in gray to indicate the specific results corresponding to this pair in the figure.
The insertion of default X2 pulses results in unexpected behavior, even degrading the performance below that of the free evolution. The reason for this outcome is elaborated in~\ref{app:x2}. 

\begin{figure}[h]
\begin{subfigure}{\columnwidth}
    \centering
    \includegraphics[scale=0.5]{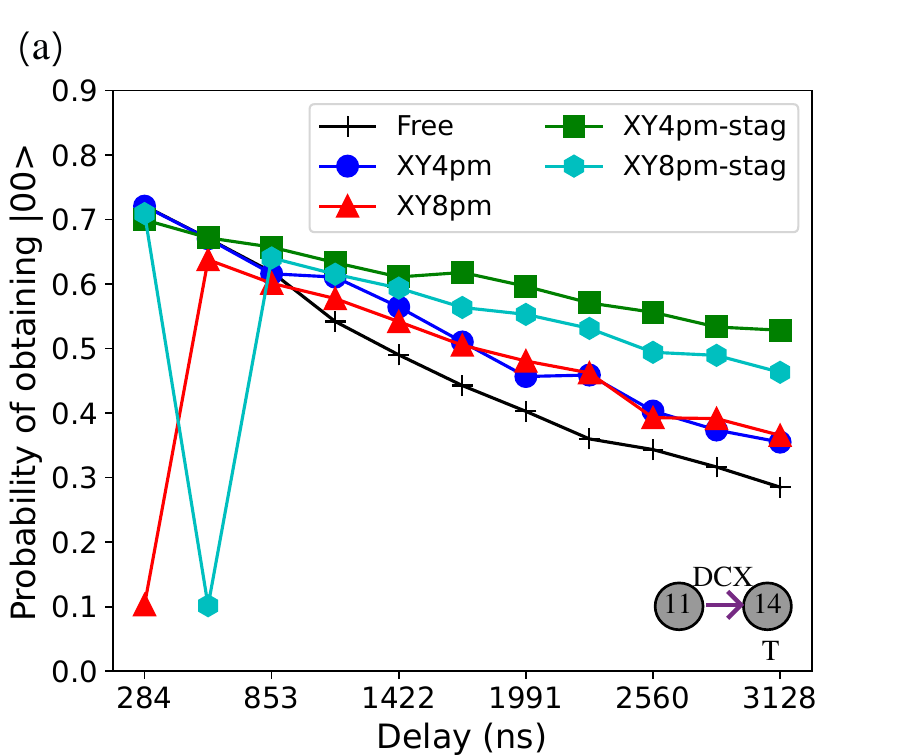}
\end{subfigure}
\begin{subfigure}{\columnwidth}
\centering
\includegraphics[scale=0.5]{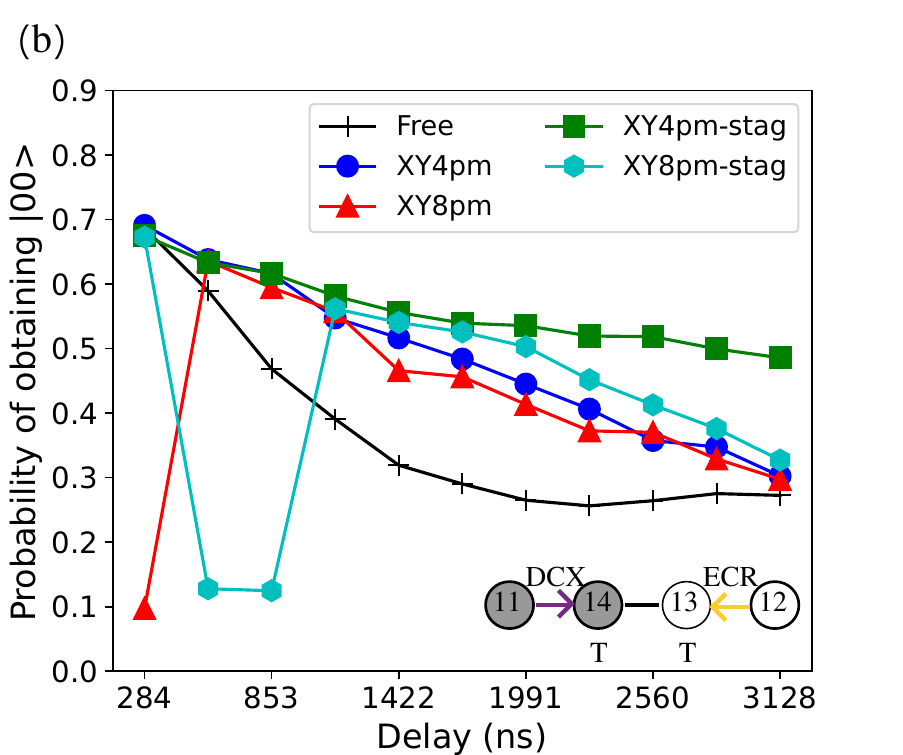}
\end{subfigure}\\

\caption{Idle-idle experiment results on qubit pairs $(q_{11}, q_{14})$ and $(q_{12}, q_{13})$ when comparing the baseline (free evolution)
with XY4pm, staggered XY4pm, XY8pm, and staggered XY8pm. (a) The probability of yielding
$|00\rangle$ for $(q_{11}, q_{14})$ when RB is applied to this qubit pair in isolation.
(b) The same probability when RB is applied to both qubit pairs simultaneously (SRB experiment). The results correspond to the gray qubit pair.}
\label{fig:idle_res_x4x8}
\end{figure}

We also demonstrate the results for XY4 and XY8 sequences in figure~\ref{fig:idle_res_x4x8}. Given the negligible difference between the results for staggered DD and its inverse, only the staggered DD is depicted. We observe ``dropouts'' in fidelity for a short idle times in the XY8 sequence which can be explained by constructive interference of higher frequency spectral noise components \cite{yuge2011noise,kotler2013noise}. In addition, the XY8 sequence shows consistently worse results than XY4, which might be due to the accumulating single-qubit gate errors. Therefore, our subsequent discussions will focus primarily on the X2 and XY4 sequences.

\subsection{Driven-idle Experiment - Randomized Benchmarking and Random SU(4)}\label{sec:cross-idle}

\begin{figure} 
\begin{subfigure}{\columnwidth}
    \centering
    \includegraphics[scale=0.5]{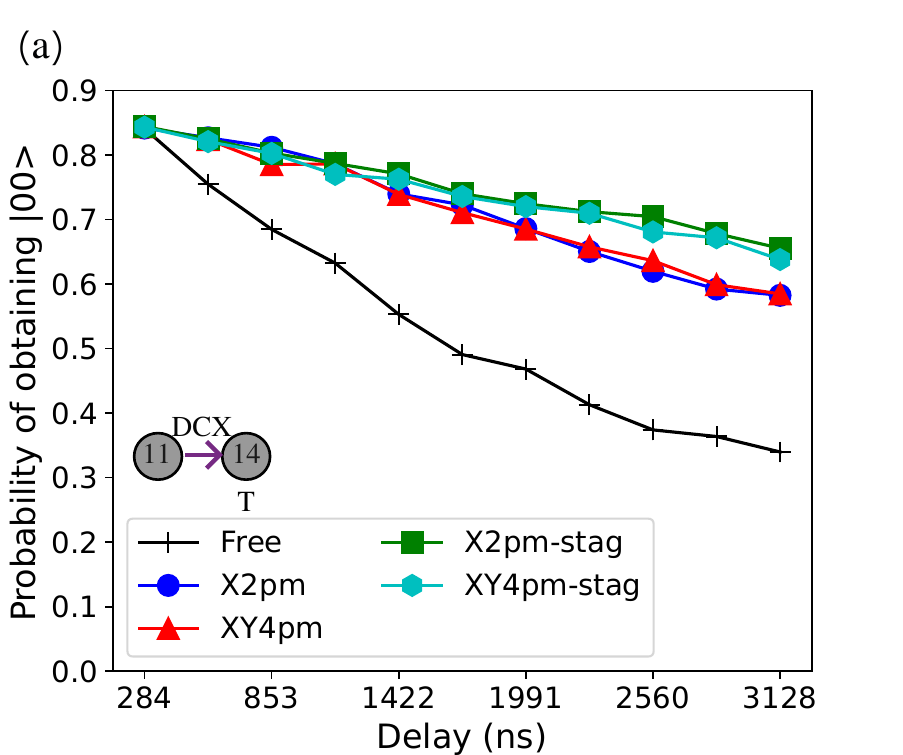}
\end{subfigure}
\begin{subfigure}{\columnwidth}
\centering
\includegraphics[scale=0.5]{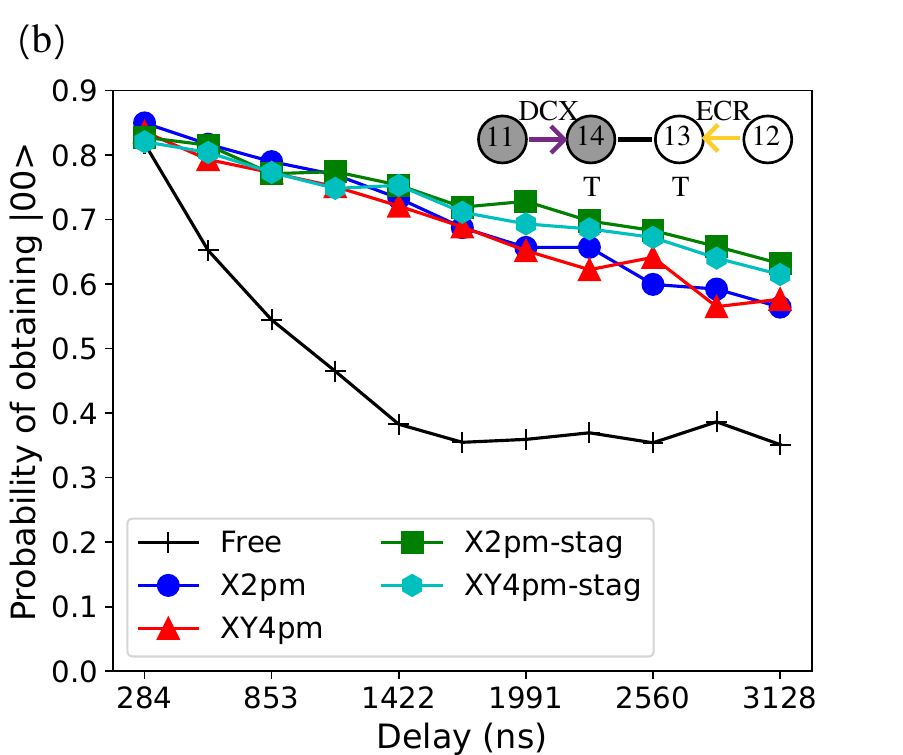}
\end{subfigure}\\

\caption{Driven-idle experiment results on qubit pairs $(q_{11}, q_{14})$ and $(q_{12}, q_{13})$. (a) The probability of yielding
$|00\rangle$ for $(q_{11}, q_{14})$ when only RB is applied. (b) The same probability when of yielding
$|00\rangle$ for $(q_{11}, q_{14})$ when RB is applied to this qubit pair while SU(4) is applied to its adjacent qubit pair $(q_{12}, q_{13})$. The results correspond to the gray qubit pair.}
\label{fig:RB_SU4}
\end{figure}

In figure~\ref{fig:RB_SU4}, we present the fidelity results when (1) applying only the RB circuit to ($q_{11}, q_{14}$) and (2) additionally performing the SU(4) circuit on the adjacent pair ($q_{12}, q_{13}$). Again, the qubits neighboring each other across the two pairs are target qubits within their respective pairs. Similar to the idle-idle experiment, the staggered DD sequences consistently outperform the non-staggered DD sequences.
In comparison to the baseline, we observe a significant enhancement in fidelity for the X2pm, XY4pm, X2pm-stag, and XY4pm-stag sequences when the RB circuit operates independently. This improvement becomes more pronounced when the RB and SU(4) circuits are executed simultaneously.

\begin{table}[h]
    \caption{\label{tab:improv}Time average fidelity for $(q_{11}, q_{14})$ when RB is applied and RB/SU(4) is applied to its neighboring qubit pairs $(q_{12}, q_{13})$. Note that, the fluctuation of RB results is due to the time difference when submitting the circuits.}
    \centering
        \begin{tabular}{c c c c c c}
        \hline
        \multicolumn{2}{c}{Experiments} & Free & X2pm & X2pm-stag & X2pm-stag-inv \\
        \hline
        \multirow{2}{*}{idle-idle} & RB & 0.65 & 0.73 & 0.88 & 0.86 \\
                                         & SRB & 0.52 & 0.71 & 0.83 & 0.85 \\
        \multirow{2}{*}{driven-idle} & RB & 0.63 & 0.85 & 0.89 & 0.88 \\
                                              & RB-SU(4) & 0.54 & 0.82 & 0.89 & 0.88 \\
        \hline
        \end{tabular}
\end{table}

\subsection{Result Analysis}

The time average fidelities $F_{\rm avg}$ for qubit pair $(q_{11}, q_{14})$, calculated using equation~\ref{eq:fid} for both idle-idle and driven-idle experiments are shown in table~\ref{tab:improv}. For these experiments, $T$ is set to \SI{3128.9}{\nano\second}. The RB results provide a baseline comparison for the two experiments. Relative to free evolution, the application of standard X2pm and the better performing variant of staggered X2pm (chosen between X2pm-stag and X2pm-stag-inv) sequences enhances $F_{\rm avg}$ by 23.4\% and 38.3\% on average. In SRB experiment, active neighboring qubit pair generates static crosstalk when it is not in the ground state, reducing the fidelity compared with the RB experiment. Nonetheless, with DD sequences, $F_{\rm avg}$ remains stable showing improvements of 36.5\% and 63.5\% for standard and staggered X2pm compared with the free evolution. For RB-SU(4) experiment, which introduces both static and driven crosstalk, leading to a decrease of $F_{\rm avg}$ during free evolution. Here, the standard and staggered X2pm sequences boost $F_{\rm avg}$ by 51.9\% and 64.8\% compared with the free evolution. Overall, the staggered (inverse) X2pm yield fidelity improvements of 12\%, 19.7\% and 8.5\% over standard X2pm across isolated RB, idle-idle (SRB), and idle-driven (RB-SU(4)) experiments, respectively. The decrease in relative effectiveness in the idle-driven experiment may be due to the twirling effect of crosstalk from random SU(4)s~\cite{Hashim2021}.

\section{Conclusion}\label{sec:conclusion}


Our proposed staggered DD protocol can be considered as an entry point for integrating DD sequences into multi-qubit quantum systems. 
Throughout the experiments, we have tailored our staggered DD protocol to fit X2, XY4, and XY8 sequences, incorporating them during unified idle times. Several advanced DD sequences, such as UDD~\cite{uhrig2007keeping}, UR~\cite{genov2017arbitrarily}, concatenated DD (CDD)~\cite{khodjasteh2005fault}, and composite pulses~\cite{counsell1985analytical}, offer opportunities for further exploration. One natural following step is to apply the staggered DD protocol to these sequences and delve into DD scheduling methods for non-unified idle time in large quantum systems. Recently, many efforts are focused on understanding noise models for quantum hardware and developing corresponding error suppression or mitigation techniques~\cite{van2023probabilistic}. An interesting direction is to synergize the staggered DD protocol with these techniques to enhance circuit fidelity further.
Moreover, standard DD sequences have been employed to protect idle qubits for quantum error correction or detection codes~\cite{krinner2022realizing,bluvstein2023logical,pokharel2023demonstration, quiroz2024dynamically}. Incorporating our staggered DD protocol into quantum error correction codes offers a pathway toward achieving fault-tolerant quantum computing. Despite the positive impact of DD sequences, challenges such as dropouts or instability across different qubits persist. Additional experiments are necessary to deepen the understanding of hardware noise sources and to devise more hardware-oriented DD sequences.

\section*{Data Availability}
The data that support the findings of this study will be openly available following an embargo at the following URL/DOI: \\
\href{https://github.com/peachnuts/CrosstalkDD}{https://github.com/peachnuts/CrosstalkDD}.

\begin{acknowledgments}

The authors acknowledge the IBM Quantum Network Hub at QuantUM Initiative of the Region Occitanie and University of Montpellier for providing additional quantum resources through its partnership with IBM. The authors would like to thank Bibek Pokharel for a careful reading of the manuscript and Ali Javadi-Abhari, Petar Jurcevic, and Christopher J. Wood for insightful discussions.

\end{acknowledgments}

\appendix
\section{Various Experimental Configurations}\label{app:1}

\begin{figure}[!h]
    \centering
    \includegraphics[scale=0.6]{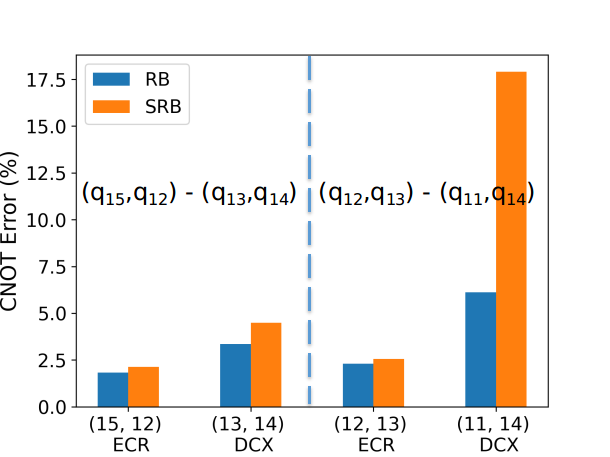}
    \caption{We perform RB and SRB on two qubit pairs and calculate the corresponding CNOT error rate. The selected qubit pairs employ either ECR or DCX type of \texttt{CX} gate. An increase in the CNOT error rate observed during SRB as compared to RB indicates the impact of crosstalk.}
    \label{fig:RB-SRB}
\end{figure}

There are two variants of $\texttt{CX}$ gates in IBM quantum hardware: Echoed Cross-Resonance CX (ECR), and Direct CX (DCX). We select several qubit pairs from \textit{ibm\_cairo} including both ECR and DCX types of CNOTs. We list the properties of these qubits,  including relaxation, decoherence, and error rate in table~\ref{tab:1}, and list the amplitudes, duration, and error rates of the qubit pairs in table~\ref{tab:2}. Note that, $\texttt{CX}$ gate is natively uni-direction in IBM quantum hardware, and the opposite direction is realized by adding extra single qubit gates. Therefore, we show explicitly the control and target qubits for each qubit pair in the table. Generally, DCX gate exhibits a lower error rate and shorter duration compared with ECR gate. However, DCX gate is more susceptible to crosstalk, as illustrated in figure~\ref{fig:RB-SRB}. Here, the CNOT error rate, derived from SRB, exhibits a notable increase compared to the error derived from independent RB for DCX, whereas the CNOT error rate for the ECR only experiences a slight increase.

\begin{figure*}[t]
    \centering
    \begin{subfigure}[b]{0.22\linewidth}
        \centering
        \includegraphics[scale=0.28]{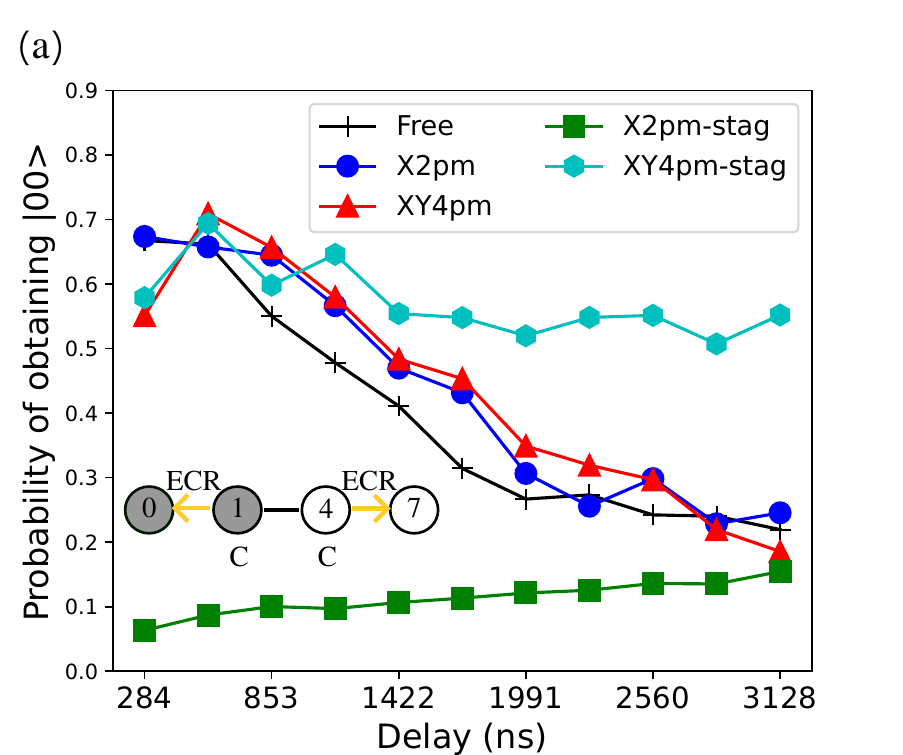}
    \end{subfigure}
    \hfil
    \begin{subfigure}[b]{0.22\linewidth}
        \centering
        \includegraphics[scale=0.28]{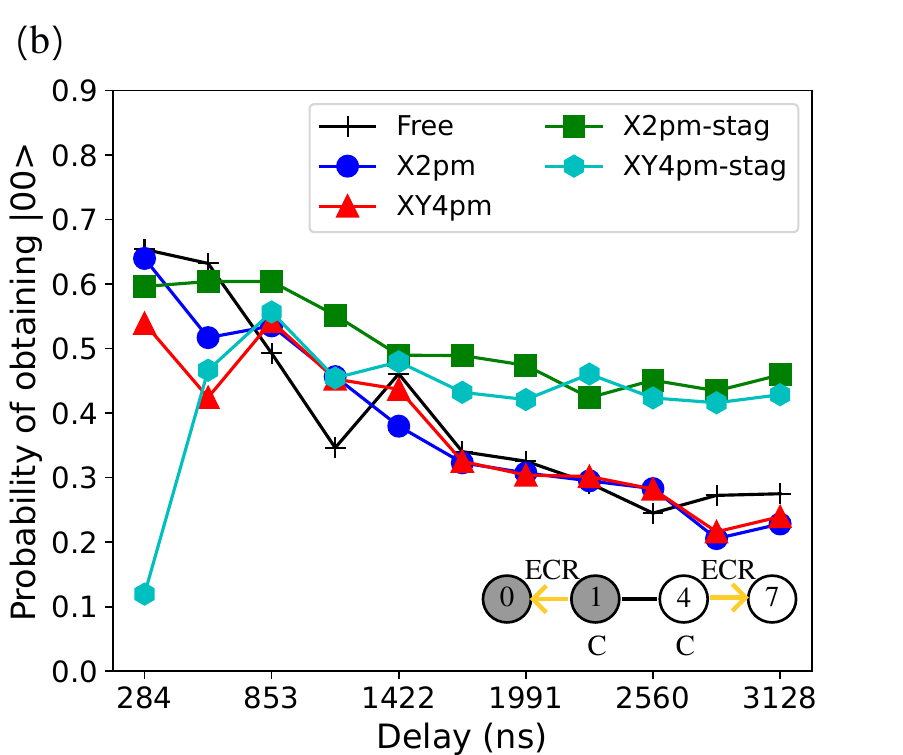}
    \end{subfigure}
    \hfil
    \begin{subfigure}[b]{0.22\linewidth}
        \centering
        \includegraphics[scale=0.28]{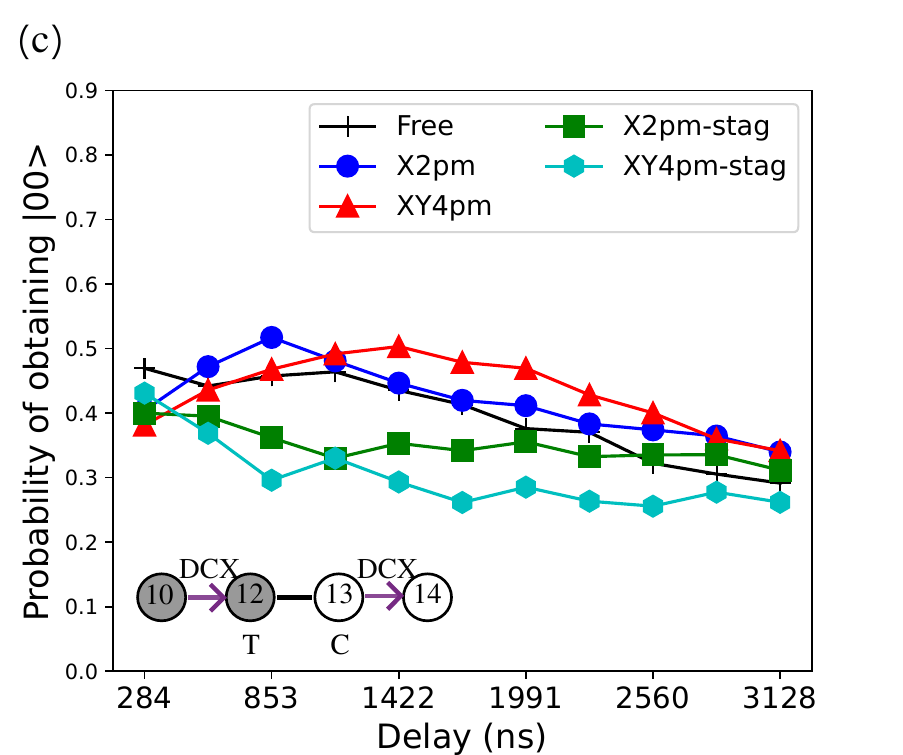}
    \end{subfigure}
    \hfil
    \begin{subfigure}[b]{0.22\linewidth}
        \centering
        \includegraphics[scale=0.28]{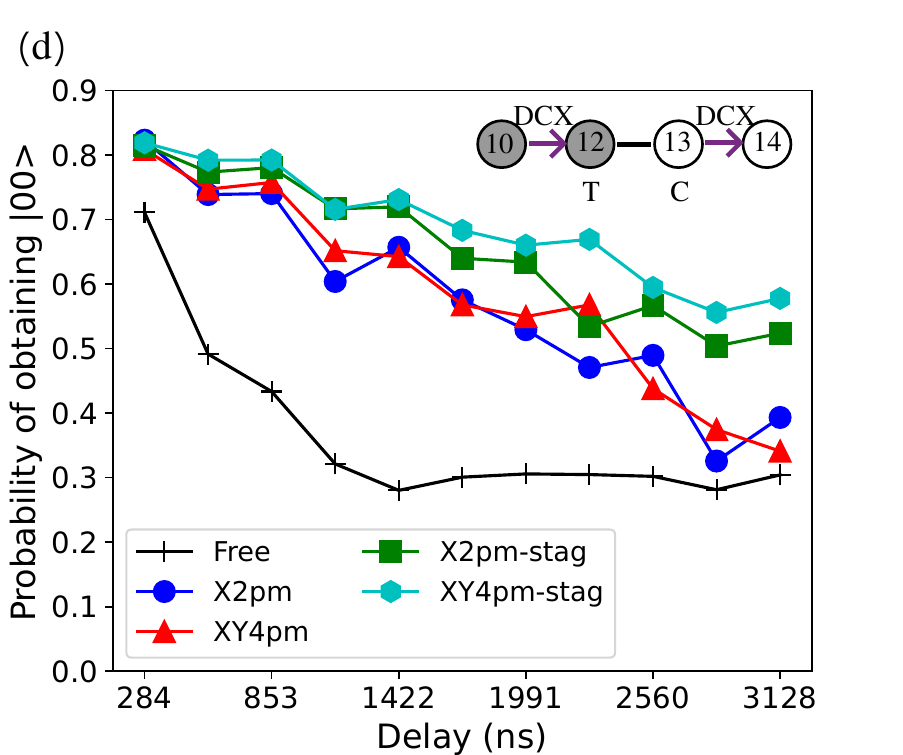}
    \end{subfigure}

    \vspace{\floatsep} 

    \begin{subfigure}[b]{0.22\linewidth}
        \centering
        \includegraphics[scale=0.28]{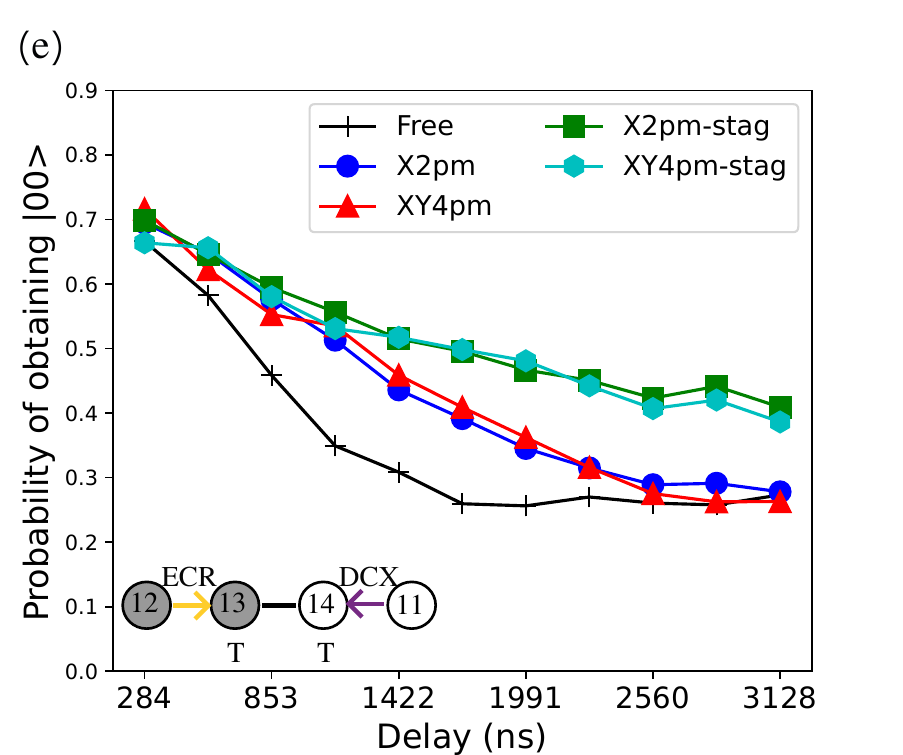}
    \end{subfigure}
    \hfil
    \begin{subfigure}[b]{0.22\linewidth}
        \centering
        \includegraphics[scale=0.28]{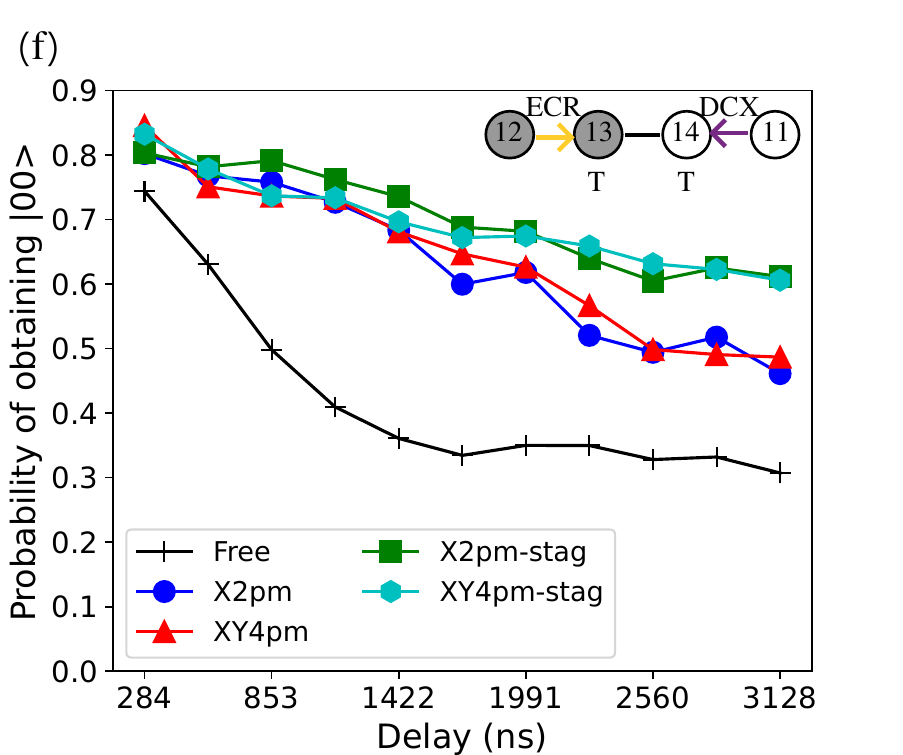}
    \end{subfigure}
    \hfil
    \begin{subfigure}[b]{0.22\linewidth}
        \centering
        \includegraphics[scale=0.28]{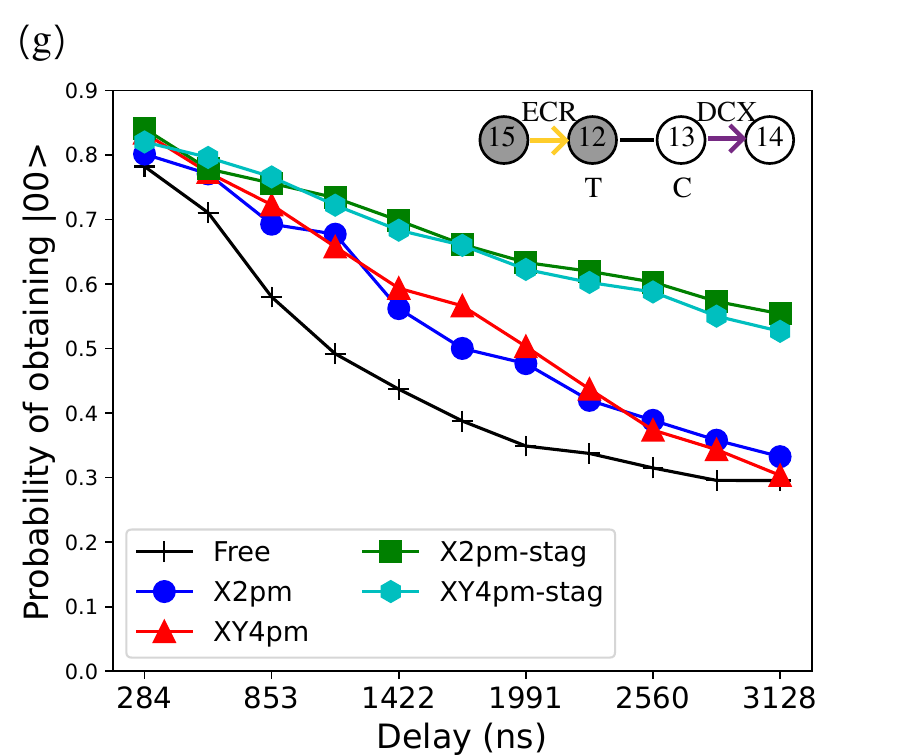}
    \end{subfigure}
    \hfil
    \begin{subfigure}[b]{0.22\linewidth}
        \centering
        \includegraphics[scale=0.28]{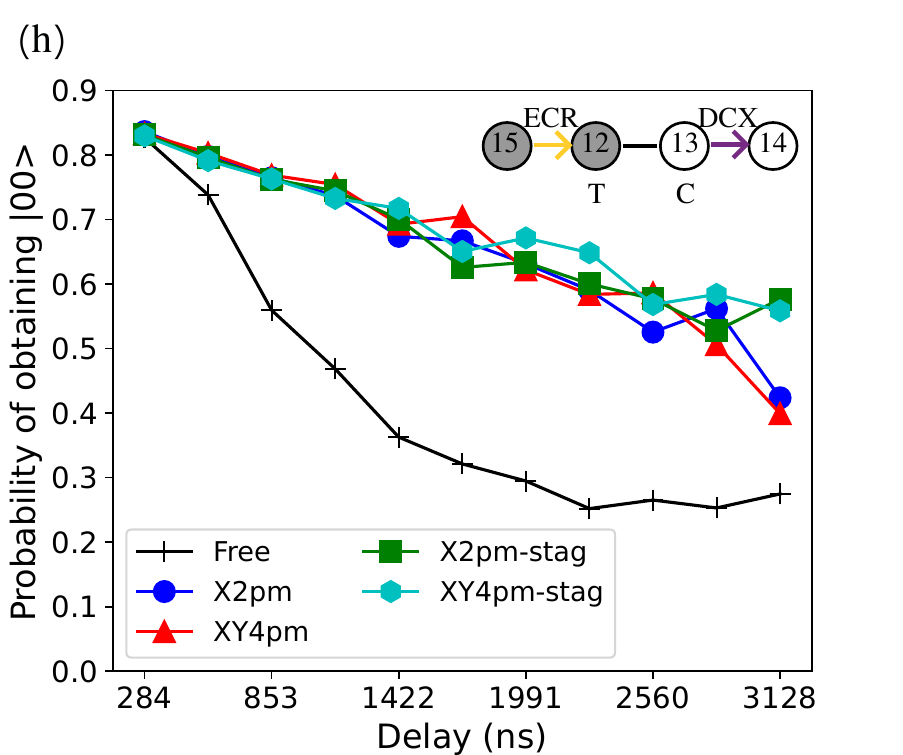}
    \end{subfigure}

   \caption{Idle-idle and driven-idle experiment results on pairs $[(q_1, q_0)- (q_4, q_7)]$} are shown in (a) and (b), on $[(q_{10}, q_{12})-(q_{13}, q_{14})]$ are shown in (c) and (d), on $[(q_{12}, q_{13})-(q_{11}, q_{14})]$ are shown in (e) and (f), and on $[(q_{15}, q_{12})-(q_{13},q_{14})]$ are shown in (g) and (h). The results correspond to the gray qubit pair.
\label{ecr-dcx-t-c}
\end{figure*}

We conduct both idle-idle and driven-idle experiments on adjacent qubit pairs across three distinct scenarios: (DCX, DCX), (ECR, DCX), and (ECR, ECR). Given the adjacency of the two pairs, one qubit from the first pair is positioned next to a qubit from the second pair, following different configurations: control-control, control-target, or target-target. The results of these experiments are displayed in figure~\ref{ecr-dcx-t-c}. In the case of driven-idle experiments, staggered DD consistently demonstrates superior performance over standard DD. However, for the idle-idle experiments, the performance of staggered DD exhibits a lack of consistency.

\begin{table}[h]
		\caption{\label{tab:CX} Single-qubit properties including relaxation ($T_1$), decoherence ($T_2$), error rate determined by randomized benchmarking, frequency, and anharmonicity.}
		\centering
  \resizebox{\columnwidth}{!}{%
		\begin{tabular}{c c c c c c}
		\hline                           
		Qubit & $T_1$ ($\mu s$)  & $T_2$ ($\mu s$) & Error rate (\%)    &  Frequency (GHz) & Anharmonicity (GHz)
	  \\ \hline

        0 & 73.69 & 144.16 & 0.02 & 5.26 & 0.32\\
        1 & 149.91 & 137.63 & 0.02 & 5.4 & 0.34\\
        4 & 71.49 & 29.85 & 0.04 & 5.19 & 0.34\\
        7 & 104.08& 68.32 & 0.08 & 5.05 & 0.34\\
        10 & 102.86& 26.56 & 0.02 & 5.23 & 0.34\\
        11 &78.89 & 9.8 & 0.04 & 5.13 & 0.34\\
        12 & 126.13& 214.87 & 0.01 & 5.11 & 0.34\\
        13 & 115.74& 148.44 & 0.02 & 5.28 & 0.34\\
        14 & 105.07& 209.43 & 0.02 & 5.04 & 0.34\\
        15 & 124.46& 164.34 & 0.01 & 4.96 & 0.34\\

    \hline
	\end{tabular}	
}
\label{tab:1}
\end{table}

\begin{table}[h]
		\caption{Two-qubit properties including type of entangling pulse for constructing $\texttt{CX}$, amplitude/duration of the pulses, error rate determined by randomized benchmarking, coupling rate $J$, and static $ZZ$ crosstalk strength.}
		\centering
  \resizebox{\columnwidth}{!}{%
		\begin{tabular}{c c c c c c c}
		\hline                        
		CX(control-target) & Type & Amplitude & Duration (dt) & Error rate (\%) & $J$ (MHz) & $ZZ$ (kHz) \\
	        \hline
         (1, 0) & ECR & 0.91 & 2400 & 2.57 & 1.54 & 40.97\\
        (4, 7) & ECR & 0.58 & 1184 & 1.64 & 2.05 & 59.04\\
        (10, 12) & DCX & 0.42 & 1008 & 0.66 & 2.06 & 56.83\\
         (11, 14) & DCX & 0.49 & 848 & 1.37 & 1.93 & 46.79\\
          (12, 13) & ECR & 0.92 & 1760 & 1.4 & 2.06 & 66.12\\
        (13, 14) & DCX & 0.33 & 992 & 0.49 & 2.11 & 103.04\\
        (15, 12) & ECR & 0.52 & 2208 & 0.9 & 1.87 & 51.19\\

    \hline
	\end{tabular}	
}
\label{tab:2}
\end{table}

\section{X2 Analysis}\label{app:x2}
\begin{figure}
    \centering
    \includegraphics[scale=0.5]{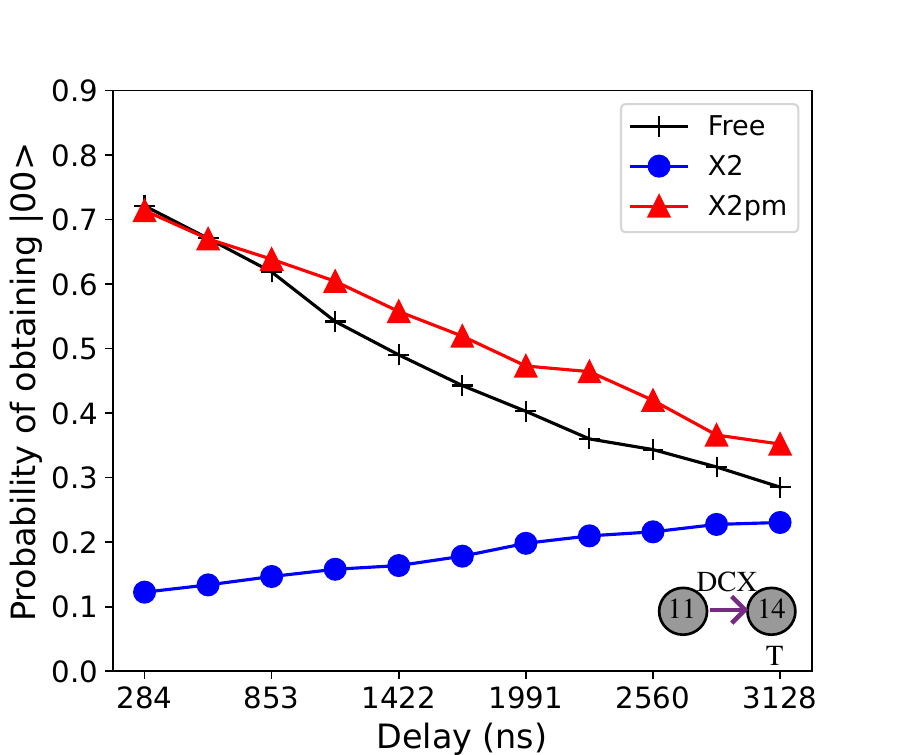}
    \caption{Comparison between free evolution, X2pm, and X2 pulse when applying RB sequences to $(q_{11}, q_{14})$.}
    \label{fig:x2-idle}
\end{figure}

The exploration of circuit fidelity for both idle-idle and driven-idle experiments reveal that the insertion of X2 sequence leads to worse outcomes than the baseline, as shown in figure~\ref{fig:x2-idle}, which is potentially due to the imperfections in the \texttt{X} gate. However, this was not reproducible in simulations with an error model informed by quantum process tomography (QPT). The QPT result on the \texttt{X} gate on \textit{ibm\_cairo} shows a high process fidelity of 0.98. Since a static $ZZ$ interaction is always present between neighboring qubits, inserting an \texttt{X} gate introduces an $IY$ component due to the non-commutation between $ZZ$ and $IX$. The second \texttt{X} gate would typically amplify this $IY$ error. However, if we apply the second \texttt{X} gate in the opposite direction, denoted as Xm, the $IY$ component is introduced in the opposite direction and tends to cancel out the previous $IY$ error.


\section{Noisy Simulation Results}\label{app:sim}
We utilize a noise simulator to perform a T2 Ramsey experiment, aiming at demonstrating the effectiveness of staggered DD in mitigating the effects of $ZZ$ crosstalk. The noise model exclusively contains thermal relaxation without any gate or readout errors. The thermal relaxation model includes both relaxation and dephasing errors. We set the induced oscillation frequency to $100$ kHz to enable more precise measurement.

The procedure begins by inserting X2pm along with its staggered counterpart into the delay of the Ramsey experiment. Within these DD-inserted circuits, we incorporate an additional $ZZ(\theta)$ gate prior to each delay, with $\theta = \omega t$, serving to simulate $ZZ$ crosstalk across time evolution $t$. We evaluate the discrepancies in detuning frequency when inserting either standard or staggered DD sequences into Ramsey experiment circuits. This comparison is performed under two conditions: with and without the introduction of $ZZ$ crosstalk.
The resulting oscillation frequencies are detailed in table~\ref{tab:3}.
In the case of no static $ZZ$, the oscillation frequency is the same as the set detuning (of 100 kHz). Static $ZZ$ is modeled by introducing $R_{ZZ}$ rotations in the idle times of the circuits, where the $Z$ rotations will be in the positive direction since the spectator qubit is in the ground state. The effect of this rotation is to change the observed oscillation frequency of the Ramsey experiment in the case of X2pm DD, whereas the staggered X2pm returns the observed oscillation frequency to nearly the set value of 100 kHz. This noise simulation outcome aligns with the theoretical analysis, validating the capability of staggered DD in canceling out $ZZ$ crosstalk.


\begin{table}[h]
		\caption{\label{tab:dd} Observed oscillation frequencies with and without $ZZ$ crosstalk for a T2 Ramsey experiment with a set detuning of 100 KHz.}
		\centering
		\begin{tabular}{c c }
		\hline                           
		Simulated model & Oscillation frequency (kHz)\\
		 \hline

        X2pm & 100 \\

        X2pm + $ZZ$ & 91.7 \\

        Staggered X2pm & 100 \\

       Staggered  X2pm + $ZZ$ & 99.1 \\
  \hline
	\end{tabular}	
 \label{tab:3}
\end{table}

\bibliography{sorsamp}

\end{document}